# Temperature effects in luminescence of associated oxygen-carbon pairs in hexagonal boron nitride under direct optical excitation within 7-1100 K range

A.S. Vokhmintsev, I.A. Weinstein

NANOTECH Centre, Ural Federal University, 19 Mira street, Ekaterinburg 620002, Russia

a.s.vokhmintsev@urfu.ru, i.a.weinstein@urfu.ru

We have studied the temperature dependencies of the photoluminescence (PL) intensity of 4.1 eV in microcrystalline powder of hexagonal boron nitride in the range of 7-1100 K. The results obtained have been analyzed within the band model of energy levels of associated donor-acceptor pairs based on impurity ($O_N C_N$) complexes. Luminescence enhancement processes at $T$<200 K and two independent channels of external thermal activation quenching are typical of the observable luminescence mechanisms under direct (4.26 eV) excitations of the samples. It has been shown that, at $T$>220 K, when directly excited, the samples diminish the PL intensity because of the processes of thermal ionization of the donor level of the $O_N$-center (122 meV) and the deep acceptor level of the $C_N$-center (1420 meV) as parts of the ($O_N C_N$)-complex. The temperature enhancement region with an activation energy of 15 meV is due to the decay of a bound Wannier-Mott exciton followed by transfer of excitation to the associated donor-acceptor pair.

Currently, much attention is being paid to the directed synthesis method, treatment and alloying of hexagonal boron nitride (h-BN) to create van der Waals heterostructures with desired properties.[1-4] This may be base for developing promising functional media for applications in nano- and optoelectronics.[5,6] In particular, owing to bandgap engineering technologies, single-photon emitters (SPEs) can be realized in h-BN, which operate in the room temperature range and exhibit zero-phonon-line energies ($E_{ZPL}$) in the visible (Vis) and ultraviolet (UV) spectral regions.[7-12] The SPE luminescence in the Vis region[7-11] is associated with complexes of point defects such as a nitrogen atom in the boron position with an anti-site nitrogen vacancy ($N_B V_N$)[9,10] and a boron vacancy with two oxygen atoms on the surface ($V_B O_2$).[8]

At the same time, upon treated with ultrasound, commercial h-BN micropowders show a bright emission of SPE in the UV region with an energy of $E_{ZPL}$=4.1 eV and several phonon replicas separated by an energy of ≈0.18 eV.[12] It is known that the indicated luminescence in the as-grown h-BN samples with impurities of carbon and oxygen is observed under subband and band-to-band excitation, and proceeds by the donor-acceptor mechanism.[13-16] Moreover, the subband excitation can be regarded as direct optical one of the ($O_N C_N$)-complex, which leads to an inter-impurity electronic transitions[16] and is characterized by the fast kinetics of luminescence decay (≈1 ns).[15,17] Earlier studies related $E_{ZPL}$=4.1 eV energy emission with radiative recombination of either "band – $C_N$-center" (carbon atom in the nitrogen position)[18,19] or "an unknown shallow donor – $C_N$-center".[15] The former is characterized by a deep acceptor level in the bandgap at a distance of 1.2-2.4 eV from the valence band top.[14,18-21] Later, the luminescence at hand was attributed to donor-acceptor recombination by a complex of associated (located at the minimum possible distance) point defects such as a nitrogen vacancy with a carbon atom in the nitrogen position ($V_NC_N$)[14,22], oxygen and carbon atoms in nitrogen positions ($O_NC_N$)[16,23], and carbon atoms in boron and nitrogen positions ($C_BC_N$).[24,25] Besides, the Ref. 26 showed that annealing of an h-BN micropowder in an $O_2$ atmosphere at a temperature of 900 °C for 2 h results in an increase in both the oxygen content from 0.25 to 0.40 wt. % and the intensity of cathodoluminescence in the 320 nm (3.9 eV) band by almost 4 times as compared to the initial sample. However, no correlation between the impurity (carbon and oxygen) concentration and $E_{ZPL}$≈4.1 eV emission intensity was found when comparing photoluminescence measurements and quantitative trace impurity analysis of variously heat-treated h-BN samples.[27] From the foregoing, it is clear that identifying the nature of the excited state of an optically active complex that luminesces with an energy of $E_{ZPL}$=4.1 eV and the role of oxygen in its formation faces some contradictions.

The objective of the present work is to investigate the temperature behavior of photoluminescence (PL) in the range of 7-1100 K for direct optical excitations of 4.1 eV luminescence centers in h-BN micropowder. The observable experimental regularities are discussed within the band model for energy levels structure of associated donor-acceptor pair (DAP) based on an impurity ($O_NC_N$)-complex.

The studied h-BN micropowder was synthesized under conditions of N deficiency, and C (2.9±0.1 at.%) and O (0.6±0.1 at.%) were the main impurities.[16,20] The PL intensity was measured in the range of 300-1100 K using a Perkin Elmer LS55 spectrometer equipped with a compact high-temperature attachment.[28,29] For taking the measurements in the range of 7-335 K, a Janis CCS-100/204N closed-cycle helium cryostat was used, combined with the recording channel of the Perkin Elmer LS55 spectrometer.[16] To exclude the possible influence of the thermally stimulated luminescence signal, the PL temperature dependencies were measured in cooling regime.[30] The cooling rates amounted to 0.1 and 1.0 K/s in the ranges of 7-335 and 300-1100 K, respectively. The emission at 3.90 eV was recorded under 4.26 eV excitation. Figure 1 presents PL spectra for the studied DAPs with $E_{ZPL}$ = 4.1 eV. The spectra exhibit obvious vibrational structure with phonon replicas of $\hbar\omega$=174 and 164 meV in the PL emission and excitation, respectively. Analyzed 3.90 and 4.26 eV bands correspond to the first phonon replicas in the PL emission and excitation, respectively (see Figure 1). This fact is consistent with the known experiments.[15,16,25,27]

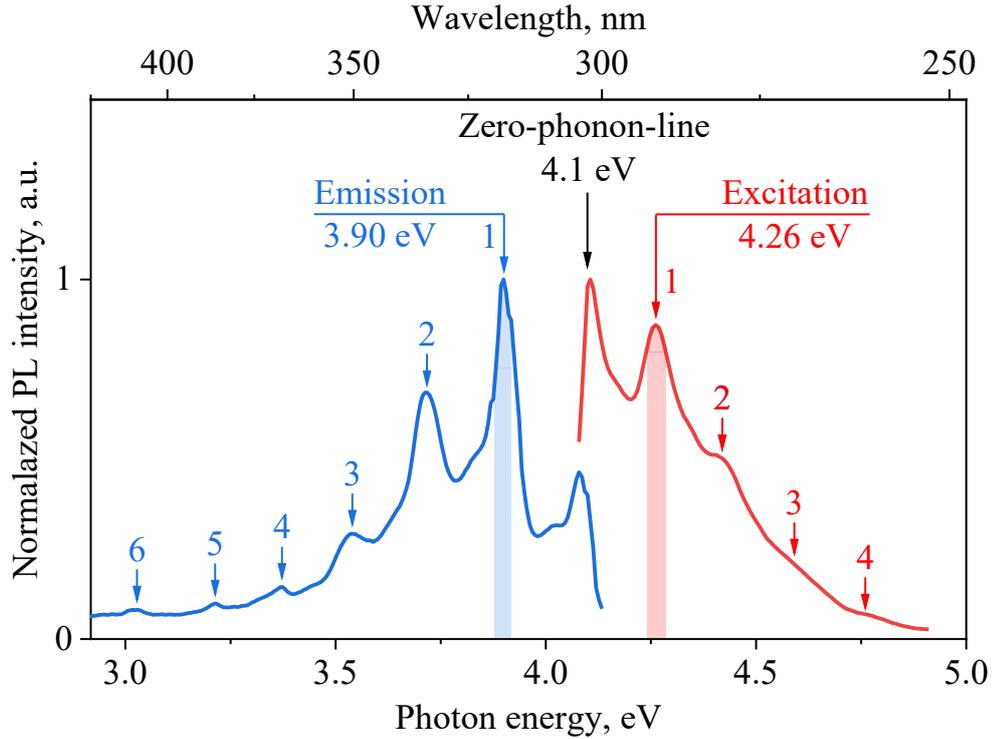

Figure 1. Normalized excitation and emission PL spectra of ($O_N C_N$)-complex measured at room temperature. Arrows show replicas positions caused by emission and excitation processes with participating the phonons with energies of $\hbar\omega$ = 174 and 164 meV, respectively.

Figure 2 presents the $I(T)$ temperature dependence measured for the PL intensity in the 3.90 eV of the h-BN powder investigated. As can be seen from the experimental curves, the luminescence intensity increases in the range of 7-220 K and reaches its maximum ($I_{max}$) in the region $T_{max}$=220-260 K. When heated to $T>T_{max}$ (≈1100 K) and excited in the band of 4.26 eV the sample loses its PL intensity up to background values. Thus, the measured dependencies of the PL response in studied h-BN are characterized by two regions with different temperature behavior, but namely, luminescence enhancement ("negative" temperature quenching) is observed at $T<T_{max}$ and the conventional thermal quenching of PL is characteristic of the region $T>T_{max}$. The dependence on Figure 2, insert, highlights two linear segments intersected at point of $(kT)^{-1}$≈12.2 eV$^{-1}$ ($T$≈950 K) in the region $(kT)^{-1}<40$ eV$^{-1}$ ($T>290$ K). In this case, it can be claimed that there are at least two non-radiative thermally activated quenching channels with energies $E_{Q1}$ and $E_{Q2}$ in the temperature range at hand.

Considering the foregoing, we conducted a further quantitative analysis of the $I(T)$ experimental dependencies within the modified Mott relation.[21,31] The latter contains independent efficiencies for two non-radiative relaxation channels (denominator) and the efficiency of the PL enhancement process (numerator)[32]:

$$I(T) = I_0 \frac{1 + p' \exp\left(-\frac{E'}{kT}\right)}{1 + p_1 \exp\left(-\frac{E_{Q1}}{kT}\right) + p_2 \exp\left(-\frac{E_{Q2}}{kT}\right)} \quad (1)$$

where $I_0$ is the PL intensity at $T = 0$ K, a.u.; $p'$ is the dimensionless pre-exponential factor of the PL enhancement process; $E'$ is the activation energy of the luminescence enhancement process, eV; $k$ is the Boltzmann constant, eV/K; $T$ is the temperature, K; $p_1$ and $p_2$ are the dimensionless pre-exponential factors of temperature quenching processes; $E_{Q1}$ and $E_{Q2}$ are activation energies of luminescence quenching processes, eV.

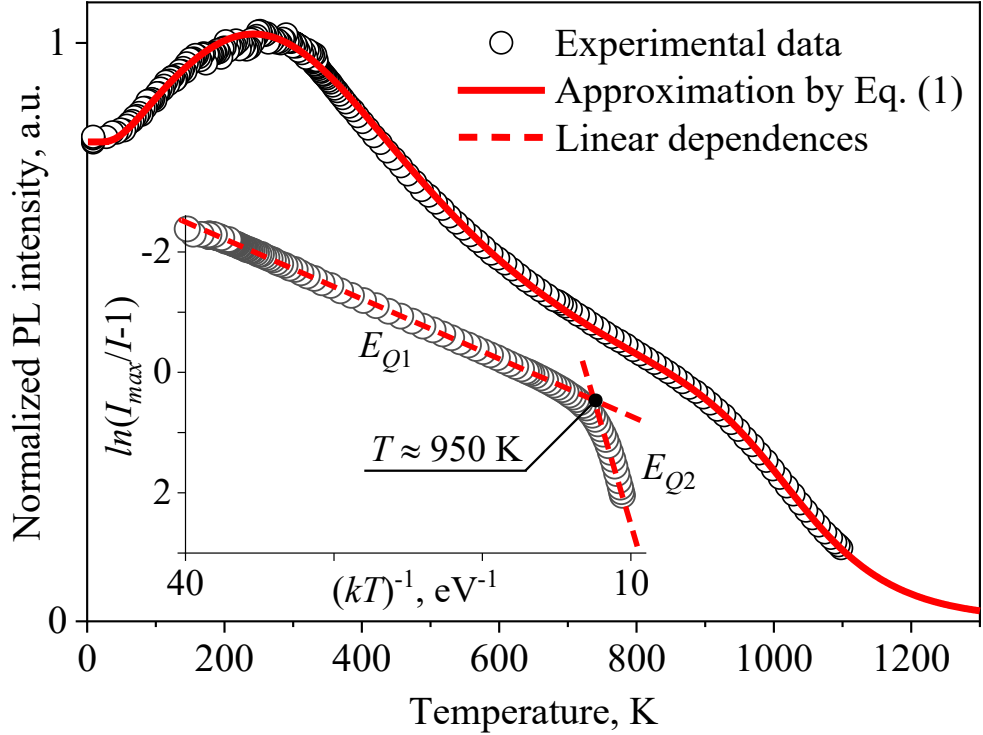

Figure 2. Normalized temperature dependencies of the PL intensity in the band of 3.90 eV at 4.26 eV excitation. Insert shows experimental data in Arrhenius coordinates.

The solid lines in Figure 2 display approximations of experimental data according to Eq. (1) with determination coefficients $R^2 > 0.998$. The following approximation parameters are obtained: $I_0 = 0.829 \pm 0.001$ a.u., $p' = 0.53 \pm 0.01$, $E' = 15 \pm 1$ meV, $p_1 = 9.0 \pm 0.1$, $E_{Q1} = 122 \pm 1$ meV, $p_2 = (2.0 \pm 0.2) \cdot 10^7$, $E_{Q2} = 1420 \pm 10$ meV.

Figure 3 illustrates a band scheme for PL processes at 3.90 eV (② transition) for the studied defect complex at excitation of the h-BN samples in at 4.26 eV (① transition). Upon absorbing a light quantum with an energy of 4.26 eV (① transition), an electron $e$ passes from the $C_N$-center level to the $O_N$-center level, generating $e$ and $h$ localized at the levels of the $O_N$- and $C_N$-centers, respectively (see Figure 3(b)). This, in turn, leads to a change in the local charge of each of the impurities and their transition to a neutral state. According to the results calculated in Ref. 33, it can be argued that the unpaired electron of the $P_z$ non-hybridized atomic orbital of the carbon atom involves in PL under study.

During selective excitation of the $(O_N C_N)$-complex, the quenching mechanism is produced by two independent channels (③ and ④ transitions) with activation energies of $E_{Q1}=122$ meV and $E_{Q2}=1420$ meV, respectively. The value of $E_{Q2}$ is consistent with the ionization energy of the deep acceptor level of the $C_N$-center in h-BN, equal to 1.2-2.4 eV.[14,18-21] It can be assumed that the value of $E_{Q1}$ corresponds to the shallow donor level of the $O_N$-center in the impurity complex under study. During direct optical excitation, the ionization of $O_N$-centers dominates due to the thermally activated transition of an electron from the first vibrational level to the conduction band.

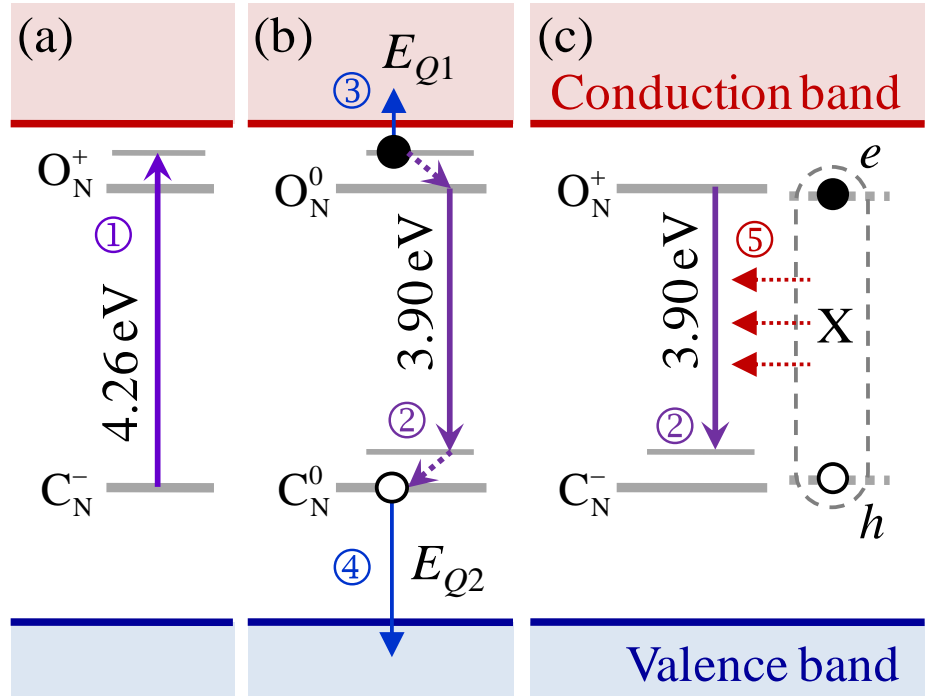

Figure 3. A band diagram of luminescence processes for an $(O_N C_N)$-complex under direct optical excitation: (a) excitation; (b) emission and external thermal quenching of PL; (c) emission and thermal enhancement of the photoluminescence.

The observable thermally activated enhancement of the PL (see Figure 2) with an energy $E'=15$ meV may be due to the formation of an exciton bound to the defect. A system consisting of closely spaced (associated) neutral donor $O_N^0$ and acceptor $C_N^0$ can be interpreted as an exciton bound with DAP (X-DAP), see Figure 3(c).[34] In this case, the X-DAP complex consists of four point-charges. Two of them are motionless: a donor ion $O_N^+$ and an acceptor ion $C_N^-$. The other two point-charges (*e* and *h*) are localized in the region of the $O_N$- and $C_N$-centers, respectively. Consequently, for the ($O_N C_N$)-complex excited in the 4.26 eV band, we can write down:

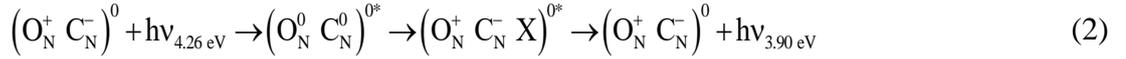

$$\left(O_N^+ \, C_N^-\right)^0 + h\nu_{4.26\,eV} \to \left(O_N^0 \, C_N^0\right)^{0*} \to \left(O_N^+ \, C_N^- \, X\right)^{0*} \to \left(O_N^+ \, C_N^-\right)^0 + h\nu_{3.90\,eV} \qquad (2)$$

According to Ref. 35, the binding energy of an exciton with an neutral impurity complex ($E_X$) satisfies the condition:

$$0.055\, E_i < E_X < 0.33\, E_i, \qquad (3)$$

where $E_i$ is the ionization energy of the impurity complex, eV.

The lower and upper limits in Eq. (3) correspond to the energy of electron detachment from a negatively charged hydrogen-like donor and dissociation of a hydrogen-like molecule, respectively. In our case, the binding energy between an exciton and the ($O_N C_N$)-complex is determined by the lowest ionization energy of the oxygen impurity ($O_N$-center) $E_i^O = E_{Q1} = 122$ meV. Using Eq. (3), we get $E_X = 7$–$40$ meV. It is seen that the activation energy of the luminescence enhancement process, equal to $E'=15$ meV, lies within the interval and matches the exciton binding energy.

In the model proposed for the X-DAP complex, the ground energy level of the bound exciton is located below the level of the excited state of the ($O_N C_N$) complex by the energy $E_X \approx E' = 15$ meV. Thus, the PL enhancement process at low temperatures of $T<T_{max}$ can be due to the decay of the exciton (see Figure 3(c), ⑤ transition) followed by excitation transfer to DAP. Hence, in the region of low temperatures, the PL of the excitons under consideration should take place. This statement is confirmed by independent studies of h-BN samples, carried out by the macro- and micro-PL method with high spectral resolution at temperatures of 8 and 10 K, respectively.[25] In Ref. 25, a low-energy shoulder shifted relative to the 4.1 eV emission line by 10-20 meV was detected for the initial and carbon-doped h-BN samples. The authors of Ref. 25 did not discuss the nature of this shoulder and the oxygen content in the samples studied.

However, according to the results of the present research work, the observable spectral shoulder can be associated with impurity-bound excitons.

Within the hydrogen-like approximation, let us evaluate the most probable radii ($r$) of the orbits for charge carriers localized near impurity ions[35]:

$$r = \frac{a_B R_y}{\varepsilon E_i}, \qquad (4)$$

where $a_B$=0.53 Å is the Bohr radius; $R_y$=13.6 eV is the Rydberg constant; $\varepsilon$ =3.5 is the relative dielectric constant for h-BN.[36] The specified value is equal to $r_O$=18.4 Å for $e$ at the $O_N$-center and $r_C$=1.5 Å for $h$ in the region of the $C_N$-center. The minimum distance between $e$ and $h$ in question exceeding the interatomic distances in the h-BN crystal lattice, it can be claimed that, at $T<T_{max}$, it is a Wannier-Mott exciton that is probably expected to be localized on the impurity ($O_N C_N$)-complex. The assumption made above is consistent with the PL enhancement experimentally recorded (see Figure 2) in the cryogenic temperature range for the h-BN samples examined.

Earlier, the research work proposed the Wannier-Mott exciton model to discuss the results of experimental studies of the optical properties of pure h-BN single crystals.[37] Later, a lot of articles have also reported on luminescence bands near the edge of the bandgap in h-BN involving impurity-bound (or trapped) excitons.[15,25,38-41]

Let us look into the process of selective excitation of the ($O_N C_N$)-complex using the model proposed and describe, step by step, the transition of $e$ between the C and O impurity levels. First, when thus excited, the $C_N$-center goes into a neutral state. In other words, $h$ localized on it emerges: $C_N^- + h\nu_{4.28\,eV} \rightarrow C_N^0 + e$. Simultaneously, a positively charged $O_N$-center resides at the smallest possible distance of $R \approx 2.5$ Å between O and C in the associated ($O_N C_N$)-pair in the nearest N positions. Due to the Coulomb repulsion, the binding energy of $h$ formed at the $C_N$-center increases by the value of $\frac{q^2}{\varepsilon R} \approx 1.64$ eV, where $q$ is the charge of an electron. Next, excited $e$ is localized at the $O_N$-center in the presence of an already neutral $C_N$-center and the impurity O passes into a neutral state: $C_N^0 + e + O_N^+ \rightarrow C_N^0 + O_N^0$. When the donor and acceptor are closely spaced to each other, the van der Waals force begins to act between them. In this case, the binding energy of $e$ localized at the $O_N$-center grows by the value of $W(R)$. We write down the expression for the radiation energy upon recombination of DAP[34,35]:

$$E_{em} = E_g - E_i^C + \frac{q^2}{\varepsilon R} - E_i^O + W(R) =$$
$$= E_g - \left(E_i^C - \frac{q^2}{\varepsilon R}\right) - \left(E_i^O - W(R)\right) \quad (5)$$

where $E_g$ is the bandgap h-BN; $E_i^O$ and $E_i^C$ are the positions of the energy levels of isolated $O_N$ and $C_N$-centers, respectively.

The Eq. (5) implies that the association of the $O_N$- and $C_N$-centers causes a shift in the position of their energy levels to the bottom of conduction band and the top of valence band, respectively. In the case of external mechanisms of temperature quenching of the PL studied, the ionization energy of the isolated $C_N$-center is equal to $E_i^C = E_{Q2} + \frac{e^2}{\varepsilon R}$ =1.42+1.64=3.06 eV. The value obtained is in agreement with the position of the 3.19 eV energy level relative to the valence band top for an isolated $C_N$-center in h-BN.[42] The calculation by the method of density functional theory within the generalized Kohn-Sham scheme also confirms the above assertion.[42]

Meanwhile, according to the Eq. (5), the $W(R)$ van der Waals summand is necessary to know to compute the ionization energy of the independent $O_N$-center. An analysis of the calculation works showed that the binding energy between two layers in h-BN varies in the range of 55-152 meV.[43,44] Suppose that, for the ($O_N C_N$)-complex at hand, the van der Waals summand takes similar values, i.e., $W(R)$=55-152 meV. In this case, taking into account the quenching energy and the position of the first phonon replica for DAP donor level of the DAP ($E_{Q1}+\hbar\omega$=122+164 meV), the ionization energy of the independent $O_N$-center is $E_i^O$=341-438 meV. The above value is consistent with an activation energy of 300-500 meV for electron trap based on $O_N$-center.[18,45]

In summary, the temperature dependencies of photoluminescence have been measured in the 3.90 eV band in the range of 7-1100 K under direct (4.26 eV) excitation of h-BN micropowder samples with C and O impurities. A region of PL enhancement in the temperature range of $T$<200 K was found. The obtained experimental data were analyzed within a model with one PL enhancement channel and two independent channels of external temperature quenching. The corresponding activation energies of the investigated processes are estimated. It is shown that with quenching of PL with energies of 122 and 1420 meV is due, respectively, to the thermal ionization of a shallow donor $O_N$-level and deep acceptor $C_N$-level as parts of the impurity ($O_N C_N$)-complex under study. Within the hydrogen-like approximation, the most probable radii of the orbits of localization of electrons and holes by the O and C impurity ions were calculated. They amount to 18.4 and 1.5 Å, respectively. The PL enhancement in the temperature dependence is assumed to be contributed by the decay of Wannier-Mott excitons, which are bound with the impurity complex in question and have a binding energy of 15 meV.

**Acknowledgments**

The work was supported by Act 211 Government of the Russian Federation, contract № 02. A03.21.0006 and by Minobrnauki research project.